\documentclass[aip,amsmath,amssymb,graphicx,reprint]{revtex4-1}

\usepackage[version=3]{mhchem}
\usepackage{balance} 
\usepackage{graphicx} %eps figures can be used instead
\usepackage{epstopdf}
\usepackage[format=plain,justification=raggedright,singlelinecheck=false,font=small,labelfont=bf,labelsep=space]{caption} 
\usepackage{fancyhdr}
\begin{document}

\title{Nanoscale spatially resolved infrared spectra from single microdroplets}

\author{Thomas M\"uller}
\altaffiliation{These authors contributed equally to this work.}
\affiliation{Department of Chemistry, University of Cambridge, Lensfield Road, Cambridge CB2 1EW, United Kingdom}

\author{Francesco Simone Ruggeri}
\altaffiliation{These authors contributed equally to this work.}
\author{Andrzej J. Kulik}
\affiliation{EPFL, Laboratory of the Physics of Living Matter, Route de la Sorge, CH-1015 Lausanne, Switzerland}

\author{Ulyana Shimanovich}
\author{Thomas O. Mason}
\author{Tuomas P. J. Knowles}
\email{tpjk2@cam.ac.uk}
\affiliation{Department of Chemistry, University of Cambridge, Lensfield Road, Cambridge CB2 1EW, United Kingdom}

\author{Giovanni Dietler}
\email{giovanni.dietler@epfl.ch}
\affiliation{EPFL, Laboratory of the Physics of Living Matter, Route de la Sorge, CH-1015 Lausanne, Switzerland}

\begin{abstract}
Droplet microfluidics has emerged as a powerful platform allowing a large number of individual reactions to be carried out in spatially distinct microcompartments. Due to their small size, however, the spectroscopic characterisation of species encapsulated in such systems remains challenging. In this paper, we demonstrate the acquisition of infrared spectra from single microdroplets containing aggregation-prone proteins. To this effect, droplets are generated in a microfluidic flow-focussing device and subsequently deposited in a square array onto a ZnSe prism using a micro stamp. After drying, the solutes present in the droplets are illuminated locally by an infrared laser through the prism, and their thermal expansion upon absorption of infrared radiation is measured with an atomic force microscopy tip, granting nanoscale resolution. Using this approach, we resolve structural differences in the amide bands of the spectra of monomeric and aggregated lysozyme from single microdroplets with picolitre volume.
\end{abstract}{}

 % Electrophoresis, Charge measurement, Molecular Biophysics
%PACS, the Physics and Astronomy Classification Scheme.


\maketitle

\section{Introduction}

Lab on a chip technologies offer a range of unique opportunities for preparation and manipulation of molecular species. In particular, the compartmentalisation of biomolecules into monodisperse, micrometer-sized droplets allows for quantitative, high-throughput biochemical studies such as directed evolution,\cite{Dittrich2005,Courtois2008,Agresti2010} screening for reagents, reaction conditions or cells,\cite{Hatakeyama2006,Hong2009,Brouzes2009} as well as for the fabrication of designer emulsions and microgels.\cite{Shah2008a,Shah2008b} Microdroplets can also allow the study of rare events, such as nucleation, and have thus enabled studies of the nucleation step of A$\beta$ aggregation\cite{Meier2009} as well as insulin amyloid growth.\cite{Knowles2011}

\balance

With the rapid development of microfluidic technologies, the need of ultra-sensitive detection methods becomes ever more pressing. A large fraction of present-day experiments rely on optical detection,\cite{Song2006,Theberge2010,Gielen2013} with alternative strategies including, for instance, electrochemistry,\cite{Luo2006,Liu2008} mass spectrometry\cite{Roman2008,Fidalgo2009,Pei2009} or Raman spectroscopy.\cite{Cristobal2006,Barnes2006} Also, infrared (IR) spectroscopy techniques have been utilised to monitor the contents of microfluidic flows.\cite{Pan2004,Chan2009} Here, we demonstrate an approach for performing off-line IR spectroscopy on the contents of single microdroplets with sub-micrometer spatial resolution.

Fourier transform infrared spectroscopy (FTIR) is a key method for studying conformational properties of proteins and in particular for inferring their secondary structure.\cite{Jackson1995,Barth2007} Exposed to IR radiation, chemical bonds undergo vibrations such as stretching, bending and rotating. In the case of proteins, this leads to a spectrum characterised by a set of absorption features in the amide bands.\cite{Jackson1995,Haris1999} Thereby, the modes most commonly used to study the structural properties of polypeptides are the amide I, amide II and amide III bands. Amide I arises mainly from \ce{C=O} stretching vibrations and is generally localised within 1690-1600~cm$^{-1}$; the exact band position is determined by the backbone conformation - in other words by the secondary structure of the protein. In contrast, amide II originates from a combination of \ce{N-H} bending and \ce{C-N} stretching and is localised around 1580-1510~cm$^{-1}$. It is still possible to associate the position of the band to the protein's secondary structure, but the fact that this band stems from a combination of two different modes makes this analysis less straightforward. Finally, the amide III band is a combination of many modes such as \ce{C-N} stretching, \ce{N-H} in-plane bending, \ce{C-C} stretching as well as \ce{C=O} bending and occurs in the range of 1300-1200~cm$^{-1}$. In practice, $\alpha$-helical structures have this band centred around 1654~cm$^{-1}$, random coil proteins show a maximum around 1640~cm$^{-1}$, and $\beta$-sheet-rich amyloidic aggregates exhibit an amide I maximum within 1610-1630~cm$^{-1}$.\cite{Sarroukh2013}

\begin{figure*}[ht]
  \centering
  \includegraphics[width=\textwidth]{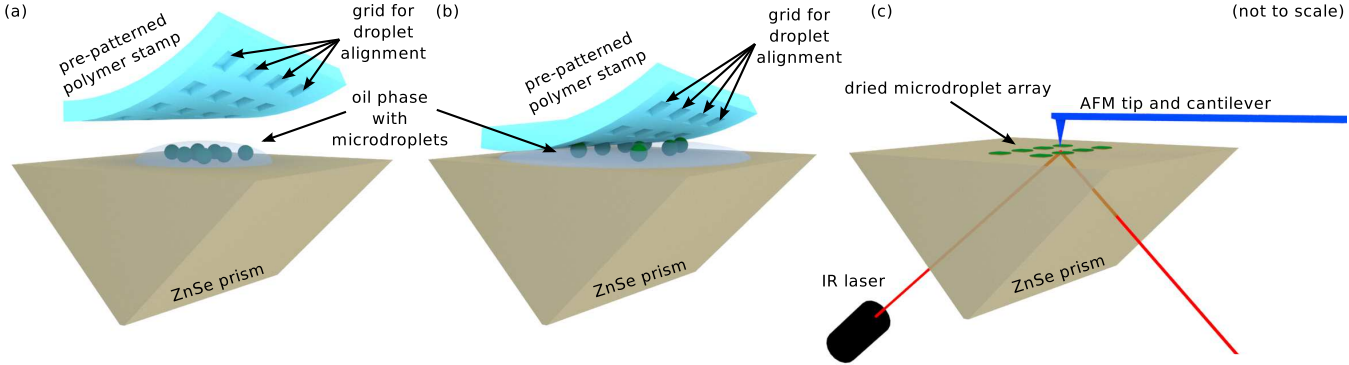}
  \caption{(a) Schematic representation of the droplets deposited on a ZnSe prism and the pre-patterned polymer stamp. The indents on the grid are $20~\mu\textrm{m}\times20~\mu\textrm{m}\times25~\mu\textrm{m}$, separated by 200~$\mu$m in each direction. (b) Alignment of the droplets on a grid by pressing the polymer onto the solution. (c) Pictogram of a laser locally heating the dried protein contents of single droplets, with an atomic force microscope measuring the resulting thermal expansion.}
  \label{fig:scheme}
\end{figure*}

To achieve sub-micron spatial resolution for protein IR spectroscopy experiments, we utilised an thermomechanical detection technique based on atomic force microscopy (AFM): if an IR pulse at a given wavelength is absorbed by a sample, the local temperature rise leads to local thermal expansion.\cite{Kjoller2010,Dazzi2010,Marcott2013} This deformation excites a mechanical resonance in the AFM cantilever which is in contact with the analyte. AFM detection of this temporary expansion of the scanned region therefore allows nanoscale resolution IR imaging and acquisition of local chemical spectra. Simultaneously with the IR-absorption image, the system is able to scan topography (with conventional contact mode) and sample stiffness (related to the frequency of the cantilever oscillations).

In order to be able to reliably locate and distinguish the protein contents of individual microdroplets for performing nanoscale IR spectroscopy as well as to provide for an enormous amount of statistics, the droplets are aligned on a ZnSe prism using a pre-patterned polymer stamp and dried overnight, as shown schematically in Fig.~\ref{fig:scheme}(a)-(c). Thereafter, spectra of single dried microdroplets containing monomeric and aggregated protein are acquired, and the ability to easily differentiate between the two demonstrates the efficacy of the presented approach.

The capability of studying the contents of single microdroplets individually with a high-precision method paves the way for a wide range of experiments harnessing the advantages of microfluidics. For instance, combination with on-chip selection techniques\cite{Ahn2006,Baroud2007,Yap2009,Franke2009,Zhang2009b} that pre-screen for a predefined species within droplets could allow for specific in-depth investigation of analytes in the context of protein aggregation as well as directed evolution. In addition, it should be emphasised that the technique of aligning individual microfluidic droplets off-chip offers further possibilities for systematic ex-situ assays beyond IR spectroscopy or AFM.

\section{Methods}
In brief, micrometer-sized droplets of protein solutions in fluorinated oil are generated via a microfluidic droplet maker and deposited on a ZnSe prism (Fig.~\ref{fig:scheme}(a)). These droplets are then aligned on a grid using a patterned stamp of polydimethylsiloxane (PDMS) as shown in Fig.~\ref{fig:scheme}(b) and dried overnight in a desiccator at room temperature or, alternatively, in an oven at $65~^\circ$C. To measure the IR spectrum, the dried protein is heated locally using a laser and the resulting thermal expansion is determined using an AFM tip, which is sketched in Fig.~\ref{fig:scheme}(c). In the following, these steps are described in more detail.

\subsection{Protein solutions}
For the monomeric solution, lysozyme from chicken egg white (Sigma-Aldrich, \textit{$\#$62970}) is dissolved in deionised water at a concentration of 6~mg/ml. Aggregates are formed by mixing 60~mg lysozyme with 1~$\%$ Sodium Azide, 200~$\mu$l of 1~M HCl, 600~$\mu$l of 10~mM HCl, 200~$\mu$l of 2~M NaCl and 5~$\mu$l of a preformed seed-fibril solution, filtering through 0.45~$\mu$m pores, followed by incubation at $65~^\circ$C for 24~h. This approach yields approximately micrometer-sized fibrils that form a gel-like structure when encapsuled as an aqueous droplet.\cite{Shimanovich2013}

\begin{figure}[htb]
  \centering
  \includegraphics[width=\columnwidth]{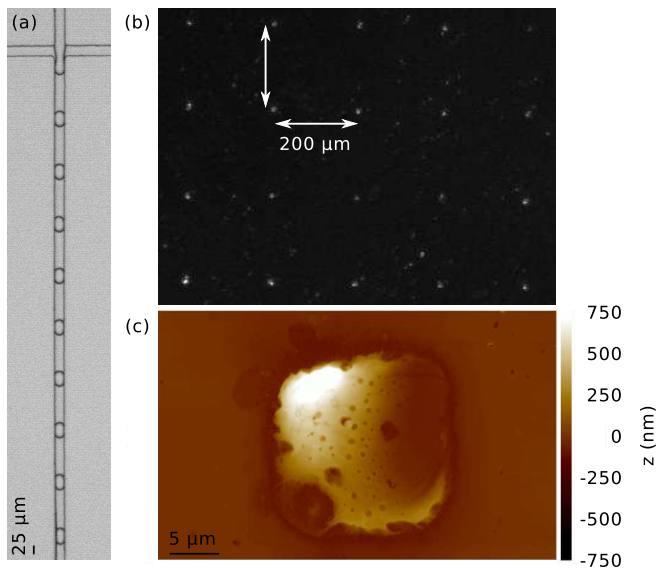}
  \caption{(a) Micrograph of droplet formation in a microfluidic device. The aqueous solution containing the protein is pushed through the central channel and encapsulated in fluorinated oil streaming in from both sides. (b) Photograph of the aligned and dried droplets after removing the polymer slab. Each bright spot is the protein content of a microdroplet, and the smears in the background stem from the fluorinated oil which has been found to not affect the IR measurements. (c) Height profile of a dried droplet - such as the ones pictured in (b) - measured by atomic force microscopy.}
  \label{fig:pics}
\end{figure}

\subsection{Droplet generation}
As depicted in Fig.~\ref{fig:pics}(a), droplets are generated using a microfluidic junction with a cross-section of $25~\mu\textrm{m}\times25~\mu\textrm{m}$, fabricated through a standard soft lithography approach.\cite{McDonald2002} The protein solution is injected through the central arm at a flow rate of 50~$\mu$l/h, whereas fluorinated oil (Fluorinert FC40, Sigma-Aldrich, \textit{$\#$F9755}) containing $2\%~w/v$ surfactant (\textit{N,Nbis(n-propyl)polyethylene oxide-bis(2-trifluoromethyl polyper fluoroethylene oxide) amide}) is pumped through the side channels at 100~$\mu$l/h. These settings result in droplets with a diameter of approximately $25~\mu\textrm{m}$ which are collected in a microcentrifuge tube.

\subsection{Droplet alignment and drying}
One ml of solution with droplets from either monomeric or aggregated protein, respectively, are pipetted onto opposite ends of the surface of an attenuated total internal reflection element prism made out of ZnSe monocrystals. Thereafter, a pre-patterned PDMS stamp is pressed onto each drop of liquid in order to align the droplets to its imprinted grid. The grid consists of  $20~\mu\textrm{m}\times20~\mu\textrm{m}\times25~\mu\textrm{m}$ indents to capture individual droplets, with a spacing of $200~\mu\textrm{m}$, and is imprinted into PDMS by soft lithography.\cite{McDonald2002} This process is illustrated in Figs.~\ref{fig:scheme}(a) and (b); for simplicity only one instance of droplet solution shown. The prism - including the alignment polymer - is then stored for 15~h in a desiccator. Alternatively, the devices can be dried at elevated temperatures - for instance, in an oven at $65~^\circ$C at ambient pressures for 15~h. This, however, will lead to further aggregation inside the droplets upon drying. Careful removal of the PDMS slab yields a regular pattern of the dried contents of single microdroplets - in the present case protein - as shown in Fig.~\ref{fig:pics}(b). AFM images of droplets dried using these two approaches are presented in the supplementary information (Sec.~\ref{sec:SI}).

\subsection{Spatially resolved infrared spectroscopy}
Samples were scanned by a commercial nano-IR microscopy system (Anasys Instruments) with a line rate between 0.02-0.08~Hz in contact mode. We used a silicon cantilever (AppNano) with a nominal radius of 10~nm and a nominal spring constant of 0.5~N/m. All images have a resolution of 512x256~pixels. Spectra were collected with a step width of $1~\textrm{cm}^{-1}$ within the range of 1200-1800$~\textrm{cm}^{-1}$ at $40\%$ of the instrument's maximal laser power. All measurements were performed at room temperature.

\section{Results and discussion}
Our results demonstrate that using the approach presented here, it is feasible to align tens to hundreds - if desired even thousands - of microdroplets, as shown in Fig.~\ref{fig:pics}(b). Such a regular deposition allows for reliable, systematic off-line assays including ultra-high precision analytical tools as, for instance, the nanoscale IR spectroscopy utilised in this work.

\begin{figure*}[ht]
\centering
  \includegraphics[width=\textwidth]{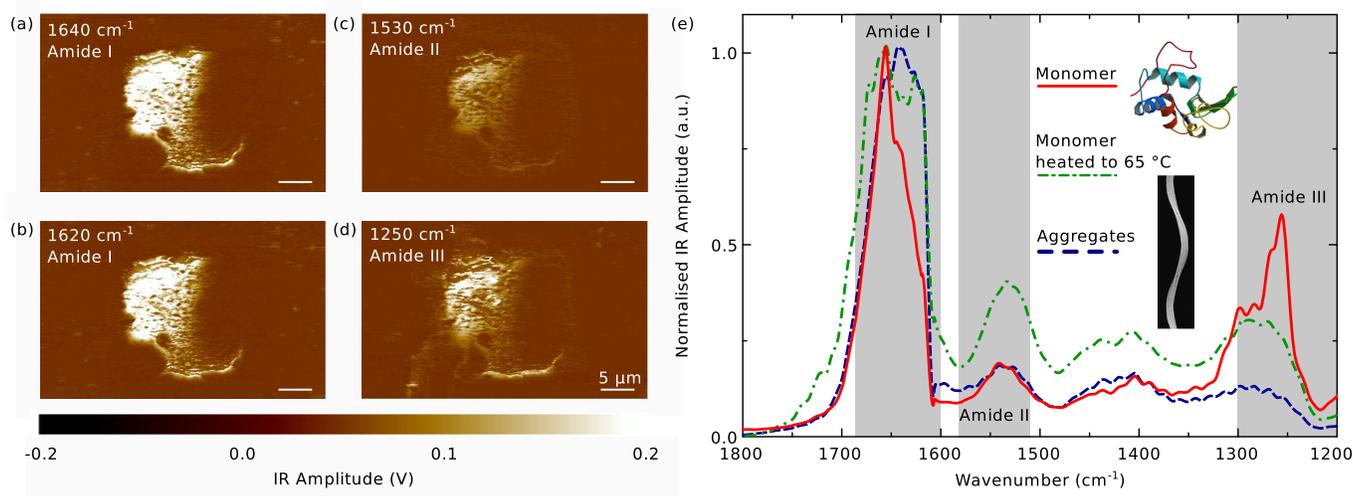}
  \caption{(a)-(d) Spatially resolved IR absorption of aggregated lysozyme from a single microdroplet at different excitation wavelengths corresponding to the amide I-III bands. (e) IR Spectra from droplets containing monomeric (solid red line) and aggregated (dashed blue line) lysozyme, as well as initially monomeric lysozyme that has been kept at $65~^\circ$C for 15~h for drying (dash-dotted green line). The measurements are averages of 12 (monomer), 21 (heated monomer) and 15 (aggregates) individual spectra taken at various locations within one dried droplet and are smoothed with a Savitzky-Golay filter. All curves are normalised to a maximal amplitude of 1. Insets for the structures of monomer and lysozyme amyloid are adapted from Refs.~\onlinecite{PDBLys} and \onlinecite{Jimenez2001}, respectively.}
  \label{fig:meas}
\end{figure*}

Figures~\ref{fig:meas}(a)-(d) show how the aggregated lysozyme from an individual microdroplet can be analysed accurately. Presented is the spatially resolved absorption of infrared radiation, determined via thermal expansion as measured by an AFM tip. The wavenumbers are fixed at 1640~cm$^{-1}$ (a), 1620~cm$^{-1}$ (b), 1530~cm$^{-1}$ (c), 1250~cm$^{-1}$ (d), corresponding to two instances from the amide I band, one from the amide II band, and one from the amide III band, respectively.

While all the four plots share their topographical features, it is readily apparent that absorption is higher in the amide I band than the amide II and III bands. Moreover, the absorption in the amide III band exhibits a stronger spatial dependence. This could be linked to the fact that this band is sensitive to different vibrational modes and therefore is influenced by local conformational changes accompanying the transition of protein from its monomeric form into aggregates.\cite{Dobson2003}

Thus, it is possible to investigate the IR absorption behaviour of the contents of individual droplets locally. Furthermore, the correlation with the height measurement from the AFM scan in Fig.~\ref{fig:pics}(c) emphasises that the recorded absorption originates from the contents of a single microdroplet.

The complete spectrum of the lysozyme aggregates in this droplet  - averaged over 12 spectra recorded at different locations and smoothed by a Savitzky-Golay filter - is given by the dashed blue line in Fig.~\ref{fig:meas}(e). As expected from the spatially resolved data, the nearly constant absorption observed in the amide I band - lightly peaked at approximately 1640~cm$^{-1}$ - is higher than in the amide II and III bands. 

Remarkably, when comparing with a spectrum taken from a droplet containing only monomeric protein (the solid red line is the smoothed average over 15 individual spectra), striking differences are apparent. First and foremost, the monomeric protein exhibits a sharp peak at around 1655~cm$^{-1}$, due to the high $\alpha$-helical content of lysozyme, and a shoulder at 1640~cm$^{-1}$ originating from random coils and $\beta$-sheets - all in good agreement with the structure of lysozyme\cite{Blake1965} and providing evidence that the secondary structural elements are largely unaffected by the gentle drying procedure. Secondly, the amide II band seems slightly shifted towards higher energies for the monomer, and thirdly, absorption in the amide III band is significantly higher in monomeric than in aggregated protein.

The shift of the amide I peak as well as the dramatic increase of absorption at 1620~cm$^{-1}$ are the typical signatures of the formation of amyloid-like cross-$\beta$ structure\cite{Sarroukh2013} and have been studied extensively for the case of lysozyme.\cite{Sassi2011} Notably, our spectra from dried lysozyme correspond very well to measurements obtained in bulk solution.\cite{Sassi2011} In fact, even if monomers already contain $\beta$-sheet domains their spectra differ from the amyloidic $\beta$-sheets and can be distinguished by the change in the location of the amide I peak.\cite{Zandomeneghi2004} Similarly, the position of the amide II band is expected to shift towards lower wavenumbers if the secondary structure changes from predominantly $\alpha$ helical to $\beta$-sheet.\cite{Jackson1995}  As the amide III band has a significantly more complex origin, the differences in the monomeric and aggregated spectra are less directly explainable, but strong deviations are reasonable bearing in mind the extensive structural modification proteins undergo during aggregation.

Finally, comparison to a spectrum from a droplet containing initially monomeric protein that was dried at $65~^\circ$C and ambient pressure for 15~h - such that aggregation can occur within the droplet - reveals that the monomeric features of the amide I and III bands are lost upon fibrillation (dash-dotted green line; average over 21 spectra). Note that due to the normalisation of the spectra to 1 the amide II band seems more pronounced. Nevertheless, the relative amplitudes of the amide II and III bands correspond very well to the spectra from the aggregates. For a ZnSe prism that was covered with SiO$_2$ and heated with droplets containing monomeric lysozyme, we observed a similar increase of absorption in the amide I band but the peak in the amide III band did not vanish, which may be an indication of partial aggregation (see Sec.~\ref{sec:SI}).

\section{Conclusions}
A technique for deposition and alignment of individual micrometer-sized droplets for their precise analysis using nanoscale spatially resolved IR spectroscopy was presented. Off-stream alignment on a grid was achieved by means of a stamp with a patterned indentation on its surface. Drying overnight fixes their content which is accessible upon removal of the polymer grid. Subsequent high-precision measurement of local IR absorption demonstrates the power of this approach to probe structural transitions in ultra small volumes. 

Spectra from droplets containing monomeric, aggregated and aggregating lysozyme were obtained and found to be readily distinguishable. In particular, the shift in the amide I band allowed us to identify an $\alpha$-to-$\beta$ secondary structure transition which is associated with amyloid formation.

While nanoscale IR spectroscopy represents a valuable analytic technique for the investigation of the contents of microfluidic droplets, the method of their alignment is not restricted to infrared spectroscopy. Indeed, any technique requiring systematic ex-situ access to the microdroplets' content is compatible with the demonstrated protocol.

We thank Pablo Aran Terol for fabricating the photolithography master for the microfluidic droplet maker and Julius Kirkegaard for help with the 3d sketches. Financial support from the Biotechnology and Biological Sciences Research Council (BBSRC), the Frances and Augustus Newman Foundation, and the Swiss National Science Foundation (SNF) is gratefully acknowledged.

\section{Supplementary Information}\label{sec:SI}
\subsection{Morphology of the dried droplets}

\footnotetext{\textit{$^{a}$~Department of Chemistry, University of Cambridge, Lensfield Road, Cambridge CB2 1EW, United Kingdom; E-mail: tpjk@cam.ac.uk}}
\footnotetext{\textit{$^{b}$~EPFL, Laboratory of the Physics of Living Matter, Route de la Sorge, CH-1015 Lausanne, Switzerland; E-mail: giovanni.dietler@epfl.ch}}
\footnotetext{\ddag~These authors contributed equally to this work.}

\begin{figure}[htb!]
  \centering
  \includegraphics[width=0.4\textwidth]{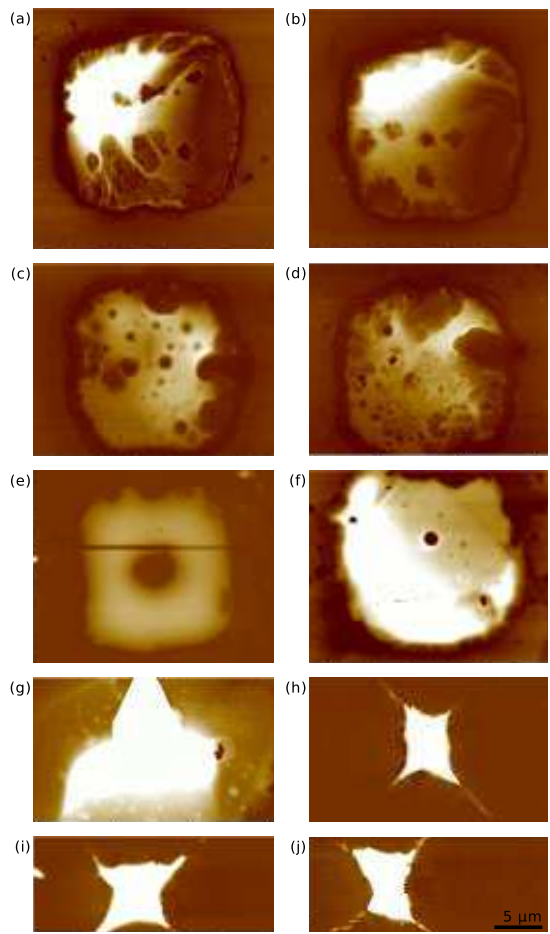}
  \caption{Atomic force microscopy images of droplets dried under different conditions. (a) Droplet containing aggregated lysozyme, deposited on a ZnSe prism and dried for 15~h at room temperature under vacuum. The full scale height is $\pm250~\textrm{nm}$. (b) Same as (a), for a different droplet. The full scale height is $\pm500~\textrm{nm}$. (c) Droplet containing aggregates, deposited on a SiO$_2$-coated ZnSe prism and dried for 15~h at room temperature under vacuum. The full scale height is $\pm500~\textrm{nm}$. (d) Same as (c) but for a different droplet. The full scale height is $\pm500~\textrm{nm}$. (e) Droplet containing monomeric lysozyme, deposited on a ZnSe prism and dried for 15~h at room temperature under vacuum. The full scale height is $\pm150~\textrm{nm}$. (f) Droplet containing monomers, deposited on a SiO$_2$-coated ZnSe prism and dried for 15~h at room temperature under vacuum. The full scale height is $\pm250~\textrm{nm}$. (g) Droplet containing aggregated lysozyme, deposited on a ZnSe prism and dried for 15~h at $65~^\circ\textrm{C}$ at ambient pressure. The full scale height is $\pm300~\textrm{nm}$. (h) Droplet containing monomeric lysozyme, deposited on a ZnSe prism and dried for 15~h at $65~^\circ\textrm{C}$ at ambient pressure. The full scale height is $\pm300~\textrm{nm}$. (i) Same as (h), but for a different droplet. The full scale height is $\pm300~\textrm{nm}$. (j) Same as (h), but for a different droplet. The full scale height is $\pm300~\textrm{nm}$.}
  \label{fig:AFMscans}
\end{figure}

Several conditions for drying and analysing the droplets were investigated. Namely, we have deposited microdroplets on pure ZnSe prisms and SiO$_2$-coated ZnSe prisms. Drying was effected by either placing the prisms under vacuum at room temperature or in an oven at $65~^\circ\textrm{C}$ at ambient pressure for a time span of 15~h. The resulting morphologies varied significantly, and are summarised in Fig.~\ref{fig:AFMscans}. Most importantly, droplets that have been heated overnight (g-j) exhibit a smaller size which may well be caused by the protein becoming more attached to the pdms stamp and therefore being removed when detaching the stamp. Note that in the main work, only data from vacuum-dried droplets is presented.

\subsection{Spectra from droplets under different conditions}

\begin{figure*}[htbp]
\centering
  \includegraphics[width=\textwidth]{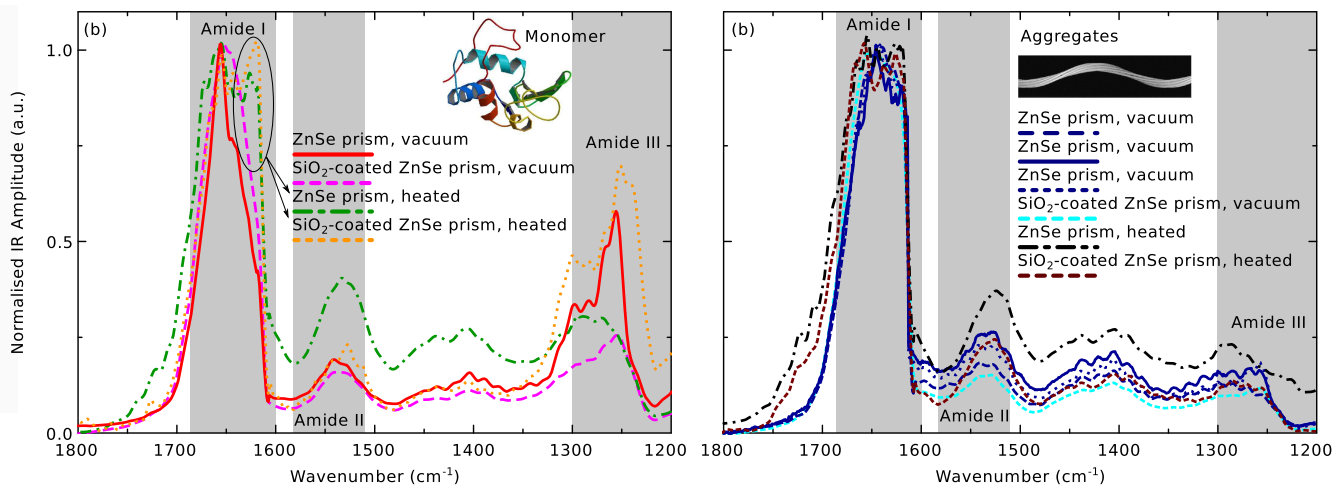}
  \caption{(a) Spectra of dried droplets containing initially monomeric protein. Shown are droplets dried in vacuum at room temperature during 15~h, deposited on a ZnSe prism (solid red line; same curve as in main text) or a SiO$_2$-coated ZnSe prism (dotted orange line), as well as dried at ambient pressure at a temperature of $65~^\circ\textrm{C}$ on a ZnSe prism (dash-dotted green line; same curve as in main text) or on a SiO$_2$-coated ZnSe prism (dashed magenta line). The black ellipse marks the appearance of a shoulder in the amide I band for the heated droplets. (b) Same as in (a) but for aggregated protein. For the vacuum-dried ZnSe prism spectra of three different droplets are presented (dashed, solid and dotted blue lines). Insets for the structures of monomer and lysozyme amyloid are adapted from Refs.~\onlinecite{PDBLys} and \onlinecite{Jimenez2001}, respectively.}
  \label{fig:spectra}
\end{figure*}

To demonstrate the repeatability of our approach, we present spectra of droplets containing monomeric and aggregated lysozyme dried under different conditions in Figs.~\ref{fig:spectra}(a) and (b), respectively. Specifically, we have investigated two different types of infrared-transparent prisms (ZnSe and SiO$_2$-coated ZnSe) and dried the droplets either at room temperature under vacuum or at $65~^\circ\textrm{C}$ at ambient pressure.

For the monomeric protein, the vacuum-dried droplets on either substrate exhibit a distinct peak around 1655~cm$^{-1}$ as well as a sharp feature in the amide III band which is pronounced more strongly for the plain ZnSe prism. Heating the droplets for 15~h to evaporate all solvents lead to a drastic increase of the absorption in the amide I band (see black ellipse in Fig.~\ref{fig:spectra}(a)) that corresponds very well to the spectra observed for aggregated protein. Furthermore, in the case of the heated, SiO$_2$-coated prism the peak in the amide III band also vanishes.

Considering the original approach for forming the aggregates, it is highly plausible that the conditions for drying the droplets at elevated temperatures are sufficient to cause at least partial aggregation of the initially monomeric protein - in particular since the evaporation of water continuously increases the protein concentration.

The spectra of the droplets containing aggregated protein presented in Fig.~\ref{fig:spectra}(b) are very similar for all conditions and demonstrate the reliability of our approach. The most notable difference is that with the additional layer of SiO$_2$ on the surface of the prism the absorption at the high-energy side of the amide I band is increased relatively to the rest of the spectrum. The peak positions, however, are unchanged. Also, the spectra from the three droplets on the same substrate (solid, dashed and dotted blue lines) indicate the quantitative variability between individual droplets with nominally the same contents.

In both cases - that is monomeric and aggregated protein - some of the differences between the spectra from ZnSe prisms and SiO$_2$-coated ZnSe prisms may be explained by the different absorption behaviour on different substrates. Indeed, in all instances of the coated prisms we observe stronger absorption towards higher wavenumbers.

Therefore, if protein aggregation from microdroplets is investigated, it is highly recommended to effect the drying under vacuum instead of at elevated temperature. Furthermore, we find that our spectra exhibit sharper features on the pure ZnSe prisms.

\subsection{Sensitivity and throughput of the method}

Given a droplet diameter of $25~\mu\textrm{m}$, the molar weight of lysozyme (14.3~kDa) and the concentration we have used for the monomers ($6~\textrm{mg/ml}\sim400~\mu\textrm{M}$), each droplet contains as little as 50~pg or 3~fmol of protein.

We can estimate the sensitivity of this approach via the minimal sample thickness (recommended is $0.1~\mu\textrm{m}$, but we have obtained spectra from samples of $h\approx60~\textrm{nm}$ height) of our Anasys nanoIR platform (http://www.anasysinstruments.com). To do so we evaluate the amount of analyte deposited on a square with length $l$ to a height $h$ using the inverse density of lysozyme ($v=0.7~\textrm{ml/g}$). This yields
\begin{equation}\label{eq:sens}
C_{min} = \frac{V_{sample}/v}{V_{droplet}}\sim\frac{h/v}{l},
\end{equation}
and with $l$ being maximally of the order of $300~\mu\textrm{m}$ the minimal concentration is around $0.2~\textrm{mg/ml}\sim20~\mu\textrm{M}$. While standard optical techniques may detect much smaller concentrations,\cite{SrisaArt2007,Guo2012} nanoIR provides a quite different set of information - namely nanometer spatially resolved IR spectra from picograms or sub-femtomoles of protein material.

The time required to obtain a spectrum of a single droplet is typically of the order of a few minutes However, performing spatially resolved measurements may take much longer. Therefore, nanoIR can by no means be considered a high-throughput technique. However, nanoIR may provide information on the contents of the droplets that is not accessible using traditional techniques and should rather be regarded as a method complimentary to on-line high-throughput techniques where measurements can be taken at kHz rates at concentrations as low as tens of nM.\cite{SrisaArt2007,Guo2012}

\footnotesize{

}

\end{document}